\documentclass[twocolumn,aps,pre]{revtex4}
\usepackage{amssymb}
\usepackage[dvips]{graphicx}
\usepackage{times}
\usepackage{color}
\newcommand{\phibar}{\overline{\phi}}

\begin{document}
\title{Local origins of volume fraction fluctuations in dense granular materials}
\author{James G. Puckett$^1$, Fr\'ed\'eric Lechenault$^{1,2}$, Karen E. Daniels$^1$}%
\email{kdaniel@ncsu.edu}
\affiliation{$^1$Department of Physics, NC State University, Raleigh, NC, 27695 USA, \\
 $^2$LCVN, UMR 5587 CNRS-UM2, Universit\'e Montpellier II, place Eug\`ene Bataillon, 34095 Montpellier, France}

\date{\today}

\begin{abstract} 
Fluctuations of the local volume fraction within granular materials have previously been observed to decrease as the system approaches jamming. We experimentally examine the role of boundary conditions and inter-particle friction $\mu$ on this relationship for a dense granular material of bidisperse particles driven under either constant volume or constant pressure. Using a radical Vorono\"i tessellation, we find the variance of the local volume fraction $\phi$ monotonically decreases as the system becomes more dense, independent of boundary condition and $\mu$. We examine the universality and origins of this trend using experiments and the recent granocentric model \cite{Clusel-2009-GMR,Corwin-2010-MRP}, modified to draw particle locations from an arbitrary distribution ${\cal P}(s)$ of neighbor distances $s$.  The mean and variance of the observed ${\cal P}(s)$ are described by a single length scale controlled by $\bar \phi$. Through the granocentric model, we observe that diverse functional forms of ${\cal P}(s)$ all produce the trend of decreasing fluctuations, but only the experimentally-observed ${\cal P}(s)$ provides quantitative agreement with the measured $\phi$ fluctuations. Thus, we find that both ${\cal P}(s)$ and ${\cal P}(\phi)$ encode similar information about the ensemble of observed packings, and are connected to each other by the local granocentric model.
\end{abstract}
\maketitle
\section{Introduction} 

Recent measurements of granular systems show that densely packed aggregates exhibit smaller fluctuations in their local volume fraction $\phi$ than more loosely-packed systems. This has been observed for several static systems \cite{Aste-2008-STG,  Briscoe-2008-EJM, SchroderTurk-2010-DSB}, as well as for a dense driven granular system \cite{Lechenault-2010-EGS} where particles with different frictional properties each exhibited the same quantitative relationship. Measurements of fluctuations in the global volume fraction $\Phi$ (or total volume) also demonstrate a similar trend in the fluctuations around a steady state value \cite{Nowak-1998-DFV, Schroter-2005-SSV, PicaCiamarra-2006-TSM,Pugnaloni-2010-SGP}. However, in several cases, the shape of the relationship between $\Phi$ and its variance did not monotonically decrease \cite{Schroter-2005-SSV, Pugnaloni-2010-SGP}. This bulk behavior may be related to the onset of cooperative effects and a possible phase transition \cite{Schroter-2007-PTS}.  In contrast, for {\itshape local} measurements of $\phi$ of static \cite{Aste-2008-STG,Briscoe-2008-EJM,SchroderTurk-2010-DSB} and dynamic \cite{Lechenault-2010-EGS} packings, the decrease in the variance was monotonic in $\phi$. The observations of this trend span experiment and simulation, various preparation protocols and particle properties, and both two and three dimensions, suggesting that a universal explanation might underly the observation. 

For dense systems, the decrease in the fluctuations of global $\Phi$ (or local $\phi$) is suggestive of a decreasing number of valid configurations as the system approaches random close packing. Heuristically, this can be understood by considering six nearest neighbor particles arranged in a ring surrounding a central particle. As the size of this ring shrinks (corresponding to increasing $\phi$ locally), the number of possible configurations for the neighbors decreases until there is a only single configuration possible in a hexagonally close packed state.

One framework in which to describe the global fluctuations in the volume fraction is the Edwards approach to the statistical mechanics of static granular systems \cite{Edwards-1989-TP}. The entropy $S(V) = k \log \Omega(V)$ increases with the number of mechanically valid, static configurations $\Omega$, which is a function of the system's volume $V$ for a constant number of particles. The change in this entropy as a function of $V$ provides a temperature-like quantity, compactivity $X \equiv \partial V / \partial S$, for which $X\rightarrow 0$ as $\Phi \rightarrow \Phi_{RCP}$ (random close packing) and $X\rightarrow \infty$ as $\Phi \rightarrow \Phi_{RLP}$ (random loose packing). The variance in either $V$ or $\Phi$ is associated with the compactivity-analog of specific heat. It remains an open question how to connect local, statistical measurements of $\phi$ to a `thermodynamics' of the bulk system for jammed systems \cite{Edwards-2003-FPS, Blumenfeld-2003-GEE}. It is even less clear how one might apply such descriptions for dynamic or even slightly unjammed configurations, as in the experiments presented here.

On the particle-scale, the local volume fraction $\phi$ is defined as the ratio of the volume occupied by the particle to the total locally-available space. One method for partitioning space is the radical Vorono\"i tessellation (also known as Laguerre cells or power diagrams), which constructs cells according to the locations and radii of the neighboring particles \cite{Richard-2001-MBA, Lechenault-2006-FVD}. Each Vorono\"i cell contains a single particle $i$, with the boundaries of the cell enclosing the set of points for which the distance $d_i$ to the particle $i$ satisfies $d_i  < d_k  + r_i^2 - r_k^2$  ($k$ are the indices of all other particles in the packing). This tessellation tiles all space, with one cell for each particle;  neighboring particles are defined as those having cells which share an edge.  The use of the radical tessellation is desirable for dense polydisperse systems in order to ensure $\phi <1$. This choice is not unique, and alternate methods for partitioning space at the particle-scale have also been utilized to similar effect \cite{Richard-2001-MBA, Ball-2002-SFG, Blumenfeld-2003-GEE, Clusel-2009-GMR}. Given a complete set of cells  which tile the volume, we can define a local volume fraction $\phi_i = v_i / V_i$, where $v_i = \pi r_i^2$ is the volume of the $i^{th}$ particle and $V_i$ is the volume of its Vorono\"i cell. The mean and variance of this distribution are denoted $\phibar$ and $\sigma^2 = \langle (\phi - \phibar)^2 \rangle$, respectively.  A bar  over a local variable indicates a mean over all particles.

For spheres in two or three dimensions, hexagonal close packed order provides the densest packing with a volume fraction of $\phi_i^{2D} = {\pi}/{\sqrt{12}} \approx 0.91 $ in two dimensions and $\phi_i^{3D} = \pi / \sqrt{18} \approx 0.74$ in three dimensions \cite{Hales-2005-PKC} . For both of these ordered packings, only a single local configuration is possible ($\phi_i = \phibar$ = const.) and the variance of the distribution ${\cal P}(\phi)$ is thus $\sigma^2=0$. In disordered systems, a jamming transition occurs at global volume fraction $\Phi_J$ and mean coordination number $\overline{z}_J$ which obey the relationship  $\overline{z}-\overline{z}_J \propto (\Phi - \Phi_J)^\alpha$ for packings above the jamming transition \cite{OHern-2002-RPF, OHern-2003-JZT}. In the presence of disorder, static packings exhibit local variations in both $z$ and $\phi$. As a result, increasing the volume fraction and coordination number of the aggregate decreases the translational and rotational randomness as well as the anisotropy \cite{Torquato-2000-RCP,SchroderTurk-2010-DSB}. Correspondingly, $\sigma^2$ is one measure of how much disorder is present in a system, and has the important advantage of not requiring knowledge of whether or not two neighboring particles are in mechanical contact. This makes it an experimentally-tractable state variable.

Due to the prevalence of the trend of decreasing $\sigma^2$ with increasing $\phibar$ in multiple jammed and unjammed experiments and simulations, we investigate how it arises. We perform experiments on an unjammed, driven system to probe the robustness of this trend with regard to boundary condition (constant pressure and constant volume) and the frictional properties of the particles (inter-particle friction coefficients $\mu~=~0.04,~0.50,$ and $0.85$). We measure $\phi$ for individual particles, and observe that $\sigma^2$ decreases linearly with increasing $\phibar$, independent of boundary condition or $\mu$. As this trend is quite similar to observations in jammed systems, we suggest that {\itshape geometry} plays an important role, rather than driving. Thus, we examine the origins of this trend via a generalization of the granocentric model \cite{Clusel-2009-GMR,Corwin-2010-MRP}, whereby we introduce randomness through the nearest neighbor distance distribution ${\cal P}(s)$.  We find that while various models of ${\cal P}(s)$ all produce the trend of decreasing fluctuations,  only the experimentally-observed ${\cal P}(s)$ provides quantitative agreement with the measured $\phi$ fluctuations. 
\section{Experiment}
\begin{figure}  
\includegraphics[width=0.95\columnwidth]{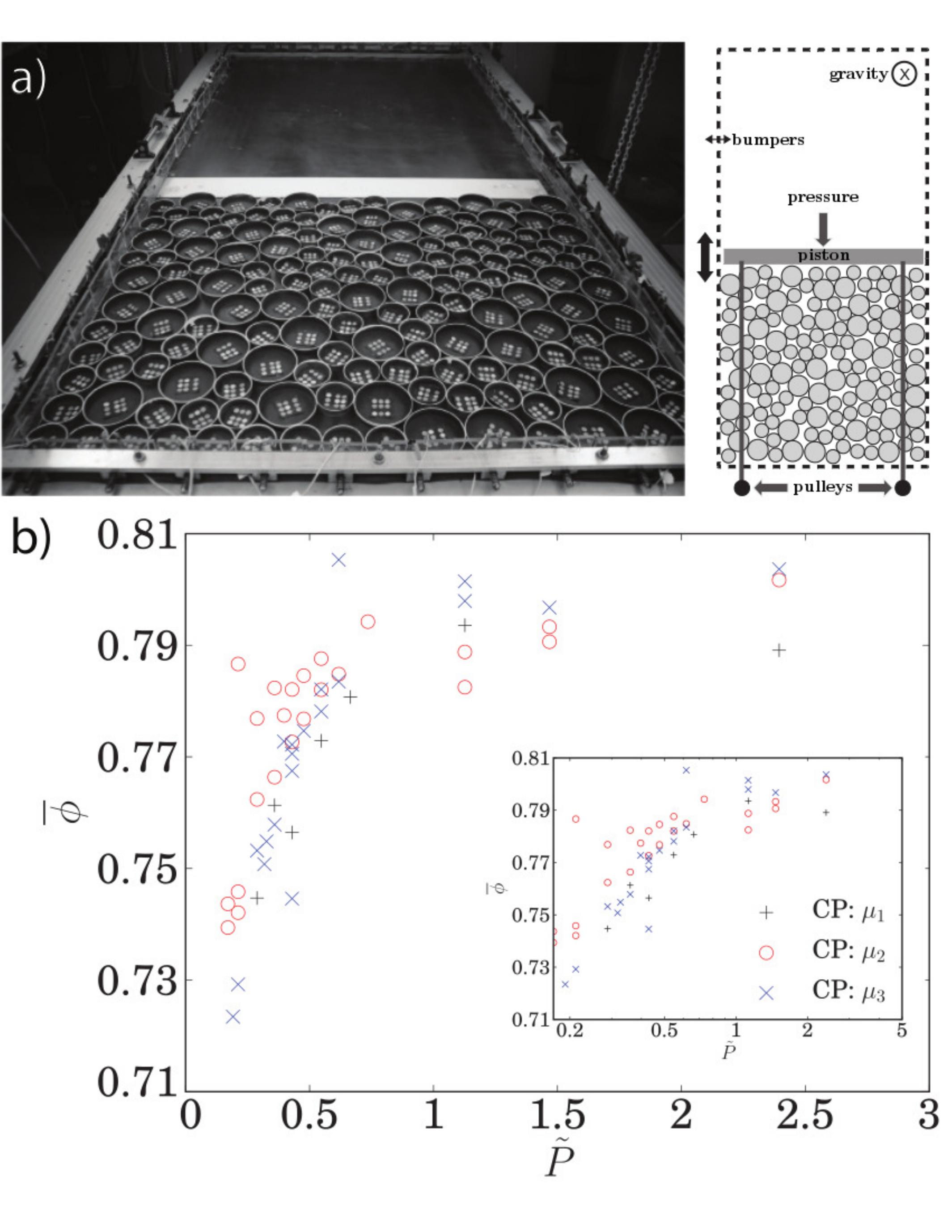}
\caption{Color online. (a) Photograph (left) and schematic (right) of the apparatus, showing confining wall in constant pressure (CP) configuration, with weights $m$ suspended from a pulley via a mono-filament line.  Constant volume (CV) is obtained by fixing the wall to the surface of the table. (b) Equation of state $\phibar(\tilde P)$ for CP experiments at $\mu_1$ ($+$), $\mu_2$ ($\circ$) and $\mu_3$ ($\times$) on linear axes and inset with semilog axes. }
\label{fig:apparatus}
\end{figure} 

The experiments are performed on a single layer of particles which are supported by an air hockey table from below and driven to rearrange by an array of sixty bumpers which form the perimeter (see Fig.~\ref{fig:apparatus}a, the same apparatus as \cite{Lechenault-2010-EGS}). The particles are a bidisperse mixture of particles with large $r_L = 43$~mm and small $r_S = 29$~mm radii with masses $m_L = 8.4$~g and $m_S = 3.4$~g respectively.  The ratio of the number of particles is fixed at 1 large particle to every 2 small particles. The particles are prepared to have one of three inter-particle frictional coefficients: $\mu_1 = 0.04$ (PTFE wrapped), $\mu_2 = 0.50$ (bare polystyrene), or $\mu_3 = 0.85$ (rubber wrapped). The air jets provide nearly frictionless contact between the particles and the base, and the table is leveled so that particles do not drift to one side in the absence of bumper driving.  While particles not experiencing collisions with either the bumpers or neighboring particles can still drift slightly due to the air jets and/or local heterogeneities in the table, these velocities are small compared with the dynamics induced by the bumpers.

Each experimental run consists of at least $10^4$ configurations captured by a camera mounted above the surface of the table. Images are collected every $2$ to $5$ seconds, according to the $\Phi$-dependent dynamical timescale of the system. The configurations are generated by the agitation from the perimeter of the packing by bumpers of width $\sim 3.2 r_S$. A pair of bumpers at the same position on opposite walls are simultaneously triggered so that no net torque is exerted on the system.  Every 0.1~s, two pairs of bumpers are randomly selected and fired, maintaining ongoing dynamics in the aggregate.

The bumpers generate evolving dense particle configurations at global volume fractions within the range $0.71 < \Phi < 0.81$. For reference, we previously measured the static random loose packing for these particles to be $\Phi_{RLP} \approx 0.81$ for $\mu_{2,3}$ \cite{Lechenault-2010-EGS}.  Experiments and simulations \cite{OHern-2002-RPF, Majmudar-2007-JTG, Lechenault-2008-CSH} with similar bidispersity have observed random close packing $\Phi_{RCP} \approx 0.84$.  The particle dynamics, driven by the bumpers at the perimeter, are caged at short time-scales and diffusive at long time-scales \cite{Lechenault-2010-EGS}.  As global $\Phi$ approaches jamming, the dynamics slow down sharply, as when the glass transition is approached in thermal systems.  The system is also well-mixing, as measured by the braiding factor \cite{Thiffeault-2010-BEP} of the trajectories growing exponentially in time \cite{Puckett-2009-GEM}. Thus, this apparatus is well-suited for generating a large number of sterically-valid (non-overlapping) configurations.

The boundary condition is determined by a movable wall on one side of an approximately 1 m $\times$ 1 m region along with bumpers on the other three sides of the aggregate. The wall, of mass $95 \, m_S$, extends the width of the table and can be configured to provide either constant pressure (CP) or constant volume (CV) boundary conditions. For CP conditions, shown in Fig.~\ref{fig:apparatus}, the wall functions like a piston and is pulled towards the aggregate by a weight of mass $m$ suspended from a low-friction pulley and a mono-filament line. For a fixed number of particles $N=186$, we perform experiments for a range of scaled pressures $\tilde P = m/m_S$ from $\tilde P = 0.17$ ($0.58$~g) to $2.41$ ($8.14$~g).  This driven granular aggregate behaves as a compressible fluid, with $\phibar(\tilde P)$ shown in Fig.~\ref{fig:apparatus}b. For CV conditions, the wall is fixed to the table so that the particles are confined within an approximately square region. We vary the number of particles $N$ from 180 to 204 (altering the global $\Phi$) while keeping the 2:1 (small:large) ratio for all $N$.  

From each image, we extract the center and radius of each particle, and perform our analysis on only the inner $20\%$ (about 40 particles), which reduces ordering effects due to the boundary \cite{Desmond-2009-RCP}. From the particle positions, we calculate $\phi$ using the {\tt Voro++} \cite{Rycroft-2006-AGF} implementation of the radical Vorono\"i tessellation. These local measurements allow us to consider the probability density function (PDF) $\cal P (\phi)$, as well as its mean $\phibar$ and variance $\sigma^2$, as a function of the three values of $\mu$ and two boundary conditions. In addition, we record the neighbors for each particle together with the inter-particle distance $s$ which separates the edges of the two particles. The distribution ${\cal P}(s)$ will provide a key input to the granocentric model.

\section{Results}  
\begin{figure}  
    \includegraphics[width=0.95\columnwidth]{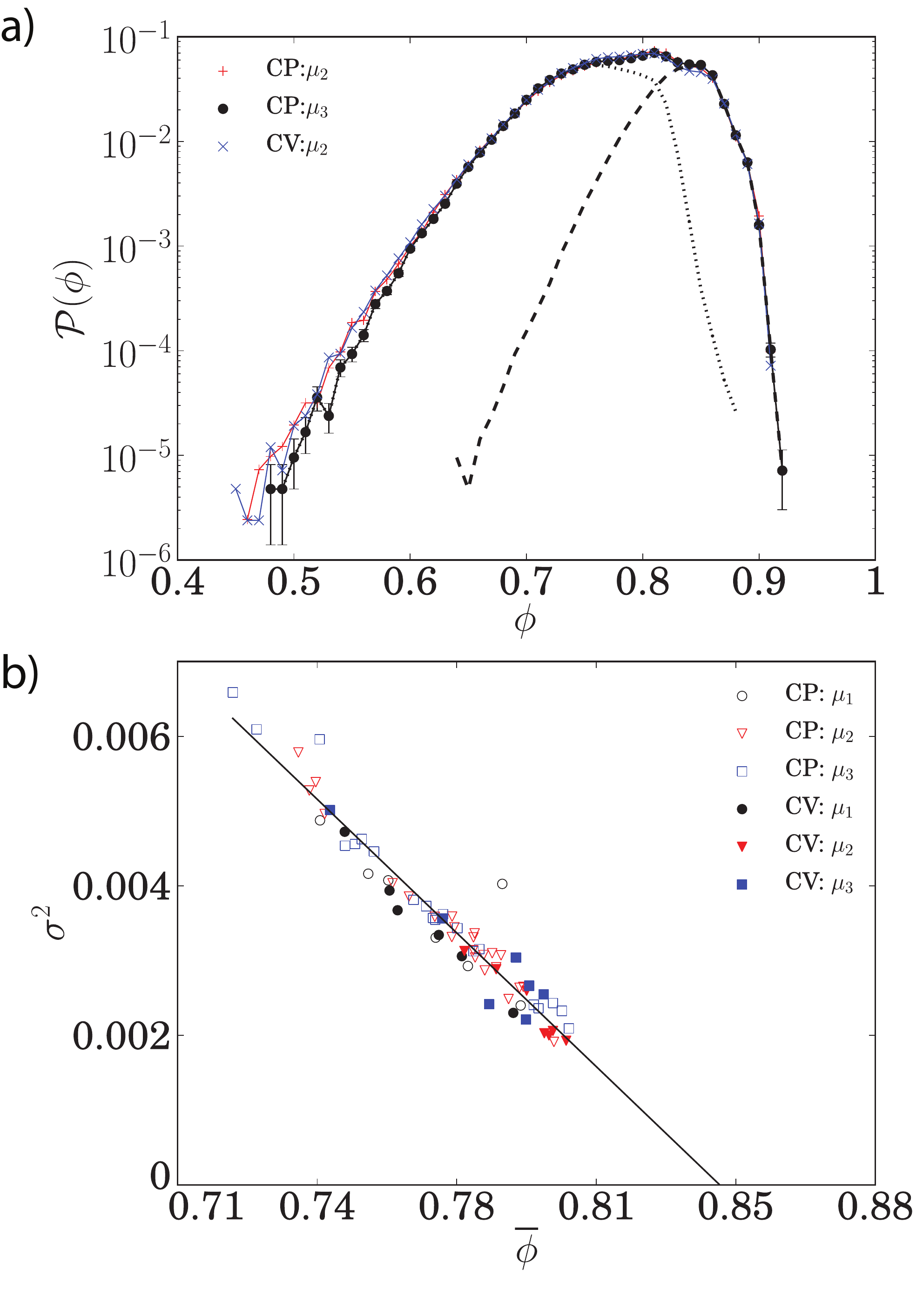}
  	\caption{Color online. (a) Representative ${\cal P}(\phi)$ for three experimental runs at $\phibar \sim 0.780$: $+$ for CP and $\mu_2$, $\bullet$ for  CP  and $\mu_3$, and $\times$ for CV and $\mu_2$.  For the run at CP and $\mu_3$, dashed and dotted lines show ${\cal P}_{L,S}(\phi)$ for large and small particles, respectively. All ${\cal P}(\phi)$, whether or not they distinguish large and small particles, are normalized by the total number of measurements, so that ${\cal P}(\phi)={\cal P}_{L}(\phi)+{\cal P}_{S}(\phi)$. (b) Mean $\phibar$ and  variance $\sigma^2$ of  ${\cal P}{(\phi)}$ measured from individual Vorono\"i cells. Each point is for a single experimental run with $\gtrsim 10^4$ configurations. Inter-particle friction is denoted by shape ($\mu_1$, $\circ$; $\mu_2$, $\triangledown $; $\mu_3$, $\square$) and boundary condition by open or filled markers (CP, open; CV, filled).  The line is a least squares fit with $\phi$-intercept at $\phi=0.842\pm0.002$.}
  	\label{fig:vphi}
\end{figure} 

As has been previously observed for {\itshape static} granular media, the ${\cal P}(\phi)$ functions arise from the inverse of a gamma-like function of the free volume \cite{Lechenault-2006-FVD, Aste-2008-EGD}. Remarkably, for experiments with similar $\phibar$, but different boundary conditions and $\mu$, we find the ${\cal P}(\phi)$ to have similar mean, variance, and shape, as shown in Fig.~\ref{fig:vphi}a. This observation typically holds at other $\phibar$ as well. As in \cite{Lechenault-2006-FVD}, the volume distribution is bimodal because large particles occupy a larger volume on average. Across all experimental runs, we observe a one-to-one relationship between the mean and variance for $\phi$ over the entire range of global $\Phi$, three values of $\mu$, and two different boundary conditions. Taking the moments $\phibar$ and  $\sigma^2$ from each distribution, we observe a single, linearly-decreasing trend, shown in Fig.~\ref{fig:vphi}b. The linear fit gives the intercept $\phi=0.842\pm0.002$ for $\sigma^2 = 0$, which is close to $\Phi_{RCP}$. Similar universality in the relationship $\sigma^2(\phibar)$  was observed in the same apparatus using a different technique for measuring $\phi$ and $\sigma^2$ \cite{Lechenault-2010-EGS}. In addition, a similar trend, perhaps with discontinuities or changes in slope, was previously observed in experimental and numerical three dimensional, monodisperse, jammed packings \cite{Aste-2008-STG, Briscoe-2008-EJM, SchroderTurk-2010-DSB}. 

\begin{figure} 
    \includegraphics[width=0.95\columnwidth]{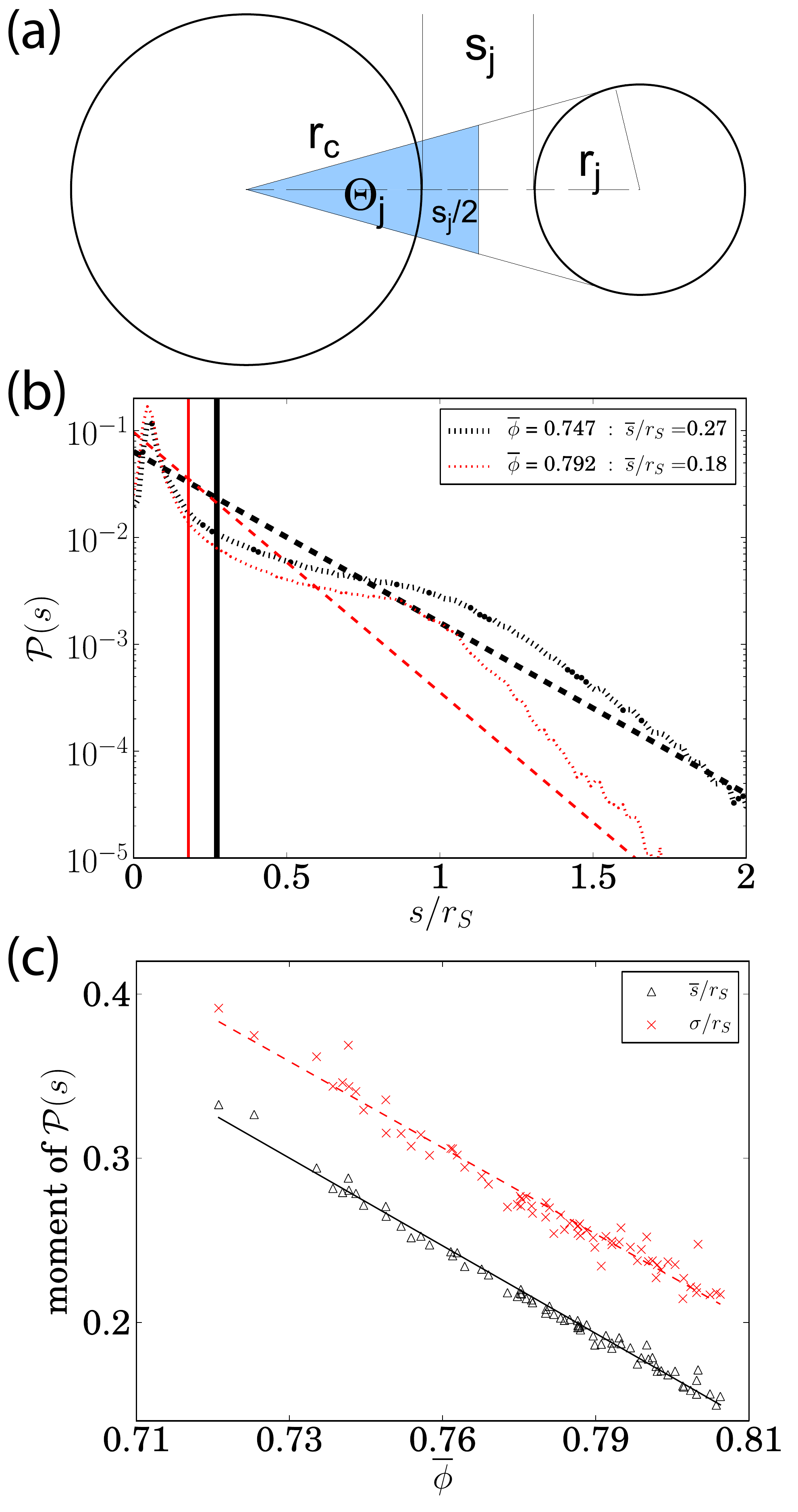}
  	\caption{Color online. 
	(a) Schematic of the central particle and one neighbor with radii $r_c$ and $r$, respectively. The shortest distance between the edges of the two particles is $s$. The shaded region is the contribution $V^*$ from this neighbor to the total radical Vorono\"i volume.
	(b) ${\cal P}^{exp}(s)$ (dotted) measured for two experimental runs at CV and $\mu_3$, with $\phibar = 0.747 $ (thick) and $\phibar = 0.792$ (thin). For comparison, exponential distribution ${\cal P}^{\lambda}(s) = \frac{1}{\lambda} e^{-s/\lambda}$ (dashed) with $\lambda \equiv \overline{s}$ and $\delta$-function distribution ${\cal P}^\delta (s) = \delta(s - \overline{s})$ (solid).
	(c) Scaled mean ($\triangle$) and standard deviation ($\times$) of $s$, for all BC and $\mu$, as a function of the local volume fraction  $\phibar$. }
	\label{fig:delta}
\end{figure}  

The $\phi$-distribution is sensitive to the location of neighbors, where each neighbor determines the boundary of one side of the Vorono\"i cell.  The probability distribution of $s$ is measured from particle edge to particle edge (see Fig.~\ref{fig:delta}a) and shown for two example runs at different $\phibar$ in Fig.~\ref{fig:delta}b. In our experiment, the shape of ${\cal P}^{exp}(s)$ is similar in all our data regardless of boundary condition or $\mu$.  After a peak in ${\cal P}^{exp}(s)$ near $s_{max} = 0.05 \, r_S$ ($1.8$ pixels), the probability falls towards a flat value until a knee at $s \approx r_S$, after which it starts to fall off exponentially with a decay set by $\overline s$.  A closer examination of ${\cal P}^{exp}(s)$ reveals that the location of $s_{max}$ is affected by whether the neighboring particles are large or small. For large-large pairs, $s_{max} \approx 0.075 \, r_S $ ($0.051 \, r_L$) and for small-small pairs $s_{max} \approx 0.026 \, r_S$, while the knee remains at $s~=~r_S$ and the exponential tail of the distribution remains unchanged. These values are independent of $\phibar$. For simplicity, we ignore the distinction between large and small particles for ${\cal P}^{exp}(s)$ with little loss in accuracy of the model results. (This choice may be inappropriate in highly polydisperse systems.)  In spite of the peculiar shape of the observed distribution, both the mean and standard deviation of $s$ appear to be smoothly set by $\phibar$, as shown in  Fig.~\ref{fig:delta}c. This suggests that a single length scale controls the distribution. 

\section{Model}  

Due to the universality of the $\sigma^2(\phibar)$ trend with regard to boundary condition and inter-particle friction, as  well as the observation of a similar relationship present for other dimensionality, polydispersity, and protocol \cite{Aste-2008-STG, Briscoe-2008-EJM, Pugnaloni-2010-SGP, Lechenault-2010-EGS}, we seek a geometric explanation for the relationship. We choose as our starting point the recent granocentric model \cite{Clusel-2009-GMR,Corwin-2010-MRP} which considers the inherently local origins of the volume fraction. The model takes the finite amount of angular space available in the vicinity of a single particle, and considers a random walk which fills this available space with particles drawn from a known distribution. We explore suitable extensions to the model and examine the implications for the universal trend in shown in Fig.~\ref{fig:vphi}b.

Three parameters govern the model: the size distribution of the particles ${\cal P}(r)$, maximum available angular space $\Theta_c$ ($2\pi$ in two dimensions, $4\pi$ in three dimensions), and fraction of $n$ neighbors in mechanical contact, $p_z \equiv z/n$. The $z$ contacts are those which provide mechanical stability for the central particle, and contribute to the $\overline{z} = 2d$ condition for isostaticity in $d$ dimensions. For measurements of the jammed emulsions which formed the inspiration for the model, $p_z \approx 0.4$ was observed and the constant separation $s$ between non-contact neighbors the free parameter used to fit the predicted ${\cal P}(\phi)$ to experiment \cite{Clusel-2009-GMR}. The model invokes randomness through both the radius distribution and $p_z$, and finds $\cal P(\phi)$ in quantitative agreement with experimental measurements.

To apply the granocentric model to un-jammed systems ($\Phi<\Phi_J$) such as this one, we consider several modifications to the inclusion of ${\cal P}(s)$. The original model draws $s$ from a binomial distribution containing $s = 0$ and a tunable $s = const.$ with probability $p_z$ and $1-p_z$, respectively. As shown in Fig~\ref{fig:delta}b, a much wider distribution of $s$ is observed in our un-jammed and driven system.  Examining ${\cal P}(s)$ for $s \approx 0$, we observe that mechanical contacts are rare; an upper bound of $p_z = 0.04$ is set by threshold of the image resolution. Therefore, we do not treat contacts and neighbors separately.

Following the original formulation of the granocentric model \cite{Clusel-2009-GMR,Corwin-2010-MRP}, there is a maximum angular space $\Theta_c$ available around a central particle with radius $r_c$ which can be occupied by neighbors.  Each neighbor of radius $r_j$ with its edge $s_j$ away from the edge of the central particle occupies an amount of space 
\begin{eqnarray}
	\Theta_j = 2 \arcsin \frac{r_j}{r_j+r_c+s_j}
\label{eqn:theta}
\end{eqnarray}
which provides a theoretical range $0 \leq \Theta_j \leq \pi$; in the experiments, only the range $0.11 \pi < \Theta_j  < 0.76 \pi$ is observed.  For a collection of $n$ randomly-selected neighbors, the total angular space occupied is  $\sum_1^n \Theta_j(r_c,r_j,s_j)$, and in two-dimensions this sum must be less than $\Theta_c = 2 \pi$. Each neighbor contributes $V_j^*$ to the Vorono\"i cell $V_i$ of the central particle, shown as the shaded region in Fig.~\ref{fig:delta}a and  given by
\begin{eqnarray}
	V_j^* &=&  \frac{r_j(r_c+ \frac{s_j}{2})^2}{\sqrt{(r_c+r_j+s_j)^2-r_j^2}}  \nonumber \\
   &=& \left(r_c+\frac{s_j}{2} \right)^2 \tan  \frac{\Theta_j}{2}
   \label{eqn:vstar}
\end{eqnarray}

To compare with the observed ${\cal P}(\phi)$, we perform a Monte Carlo simulation which draws particles from the 2:1 bidisperse size distribution and $s_j$ from a specified ${\cal P}(s)$. For each $r_c$, neighbors are sequentially selected at random from these two distributions. For each neighbor,  we calculate the $\Theta$-contribution  according to Eq.~\ref{eqn:theta}. The random process continues until $\sum_1^{n+1} \Theta_j>2\pi$, at which point insufficient angular spaces available for the last randomly-selected particle. Only the $n$ neighbors are retained and used to calculated $V_i=\sum_1^n{V_j^*}$ for the central particle. This process is repeated for $10^4$ different seeds in order to obtain a distribution of local volume fractions $\phi_i = \pi r_c^2 / V_i^*$.
 
In calculating $V_j^*$, we make one additional adjustment to account for the fact that the rejection of the $n+1$ neighbor leaves behind a neighbor-less gap of size $\Theta_{ex} \equiv 2\pi - \sum_1^n \Theta_j$. Failure to account for this gap leads to an overestimation of $\phi$, which we correct by apportioning $\Theta_{ex}$ among the neighbors in proportion to the angular space they already occupy. The adjusted angle occupied by each neighbor becomes
\begin{eqnarray}
	\tilde{\Theta}_j &=& \Theta_j \left(1 + \frac{\Theta_{ex}}{\Theta_c} \right)
	\label{eqn:thetarescale}
\end{eqnarray}
so that $\Theta_c = \sum_1^n \tilde{\Theta}_j = 2\pi$. Using this adjusted value, Eq.~\ref{eqn:vstar} becomes
\begin{eqnarray}
	\tilde{V}_j^* &=&  \left( r_c+\frac{s_j}{2} \right)^2 \tan \frac{\tilde{\Theta}_j}{2} 
	\label{eqn:vstarrescale}
\end{eqnarray}
and $\phi \equiv \pi r_c^2 / \sum_1^n{{\tilde V}_j}$ provides a better model of the local Vorono\"i volume, as the angular space surrounding the particle is now completely occupied by neighbors.  Note that the construction of the boundary at the half-distance ($r_c+s_j/2$) between the edges of the particles does not result in a cell with realistic Vorono\"i shape. This is also true for a boundary drawn at a more Vorono\"i-like distance $\frac{1}{2}( (r_c+r_j+s_j)^2+r_c^2-r_j^2) / (r_c+r_j+s_j)$ from the central particle. In either case, we find that the model reproduces the observed $\bar \phi$, but the half-distance construction quantitatively predicts $\sigma^2(\phi)$ better than the Vorono\"i-like construction.  Therefore, we use the half-distance construction for the granocentric model in the results that follow.
\begin{figure} 
    \includegraphics[width=0.95\columnwidth]{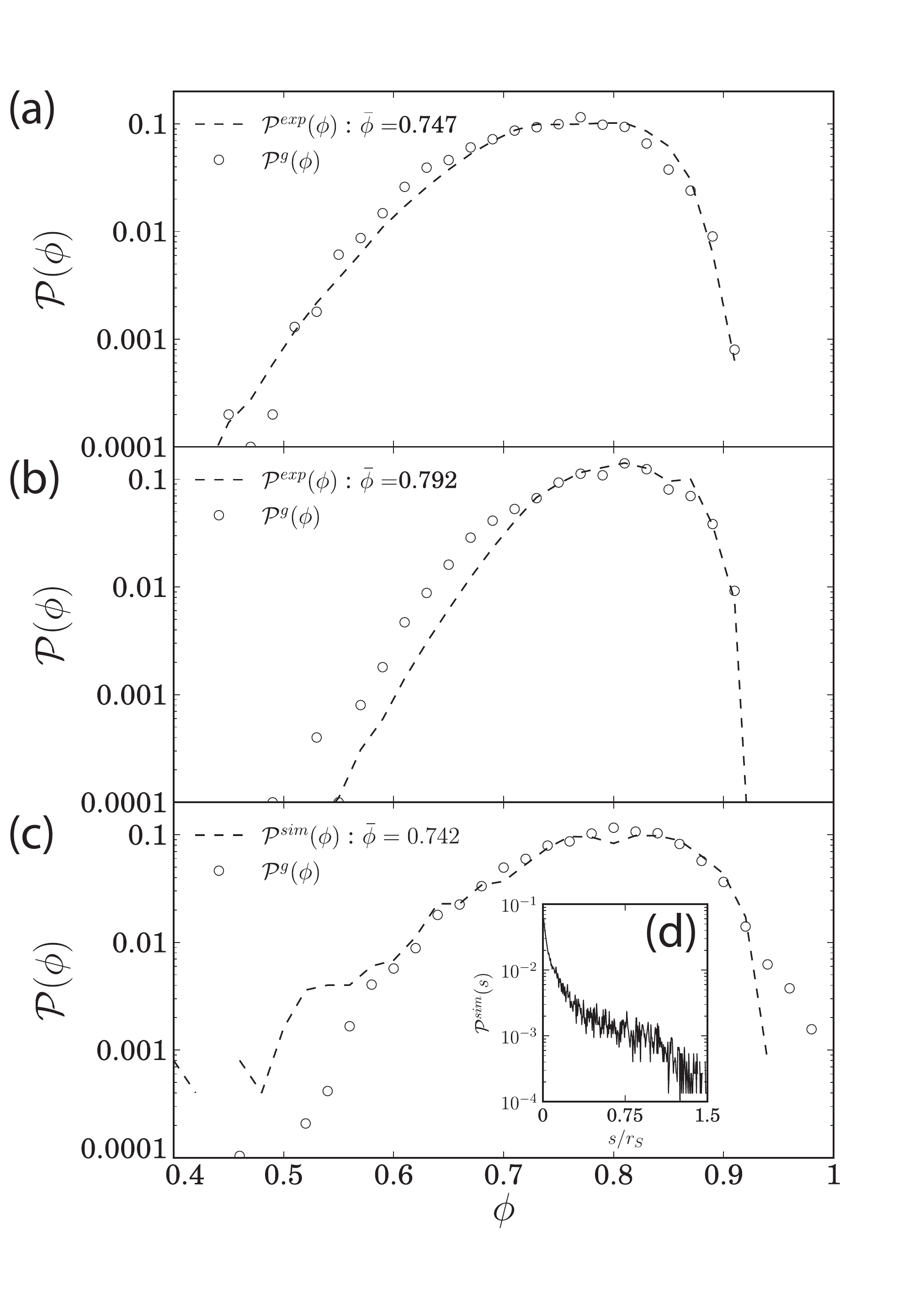}
  	\caption{(a,b,c) Measured ${\cal P}(\phi)$ (dashed line) and calculated granocentric prediction ${\cal P}^{g}(\phi)$ ($\circ$).  The granocentric prediction is calculated using (a,b)  ${\cal P}^{exp}(s)$ from Fig.~\ref{fig:delta}b and (c) ${\cal P}^{sim}(s)$ for a simulated packing with $N=10^4$ particles shown in (d).}
	\label{fig:pdf-phi}
\end{figure} 

We start from four different ${\cal P}(s)$ distributions: (1) the experimentally-measured ${\cal P}^{exp}(s)$; (2) ${\cal P}^{exp,cut}(s)={\cal P}^{exp}(s<r_S)$, the experimentally-measured distribution without the low-probability knee; (3) an exponential distribution ${\cal P}^{\lambda}(s)=\frac{1}{\lambda} e^{-s/ \lambda}$ with $\lambda \equiv \overline{ s }$ and (4) a delta function ${\cal P}^\delta(s) = \delta(s - \overline{s})$.  The effect of the low probability but large-$s$ tail of ${\cal P}^{exp}(s)$ is illustrated by ${\cal P}^{exp,cut}(s)$. While ${\cal P}^{\lambda}(s)$ and ${\cal P}^{\delta}(s)$ may not be physically realistic, these artificial distributions are chosen to show the strong dependence of ${\cal P}(\phi)$ on the functional form of ${\cal P}(s)$. Each of the four distributions is shown in  Fig.~\ref{fig:delta}b for comparison. In order to examine the ${\cal P}^g(\phi)$  which arise from these four $P(s)$, a Monte Carlo process draws a random $r_c$ to form the basis for the $i^{th}$ cell of the distribution. Around that central particle, sequential random neighbors of size $r_j$ are placed at random separations $s_j$ until the available angular space is used up; this collection of neighbors provides a value $\phi_i$ for that cell. This same Monte Carlo method is repeated to generate $10^4$ values of $\phi_i$ and thereby compute ${\cal P}^g(\phi)$ for of the four ${\cal P}(s)$ for each experimental run.
Two advantages of a Monte Carlo simulation over the semi-analytical techniques of \cite{Clusel-2009-GMR,Corwin-2010-MRP} are that it permits the use of the experimentally-measured ${\cal P}^{exp}(s)$ as an input to the model, in addition to the ability to redistribute $\Theta_{ex}$ among the randomly selected neighbors.

\section{Comparison}  

\begin{figure}
    \includegraphics[width=0.95\columnwidth]{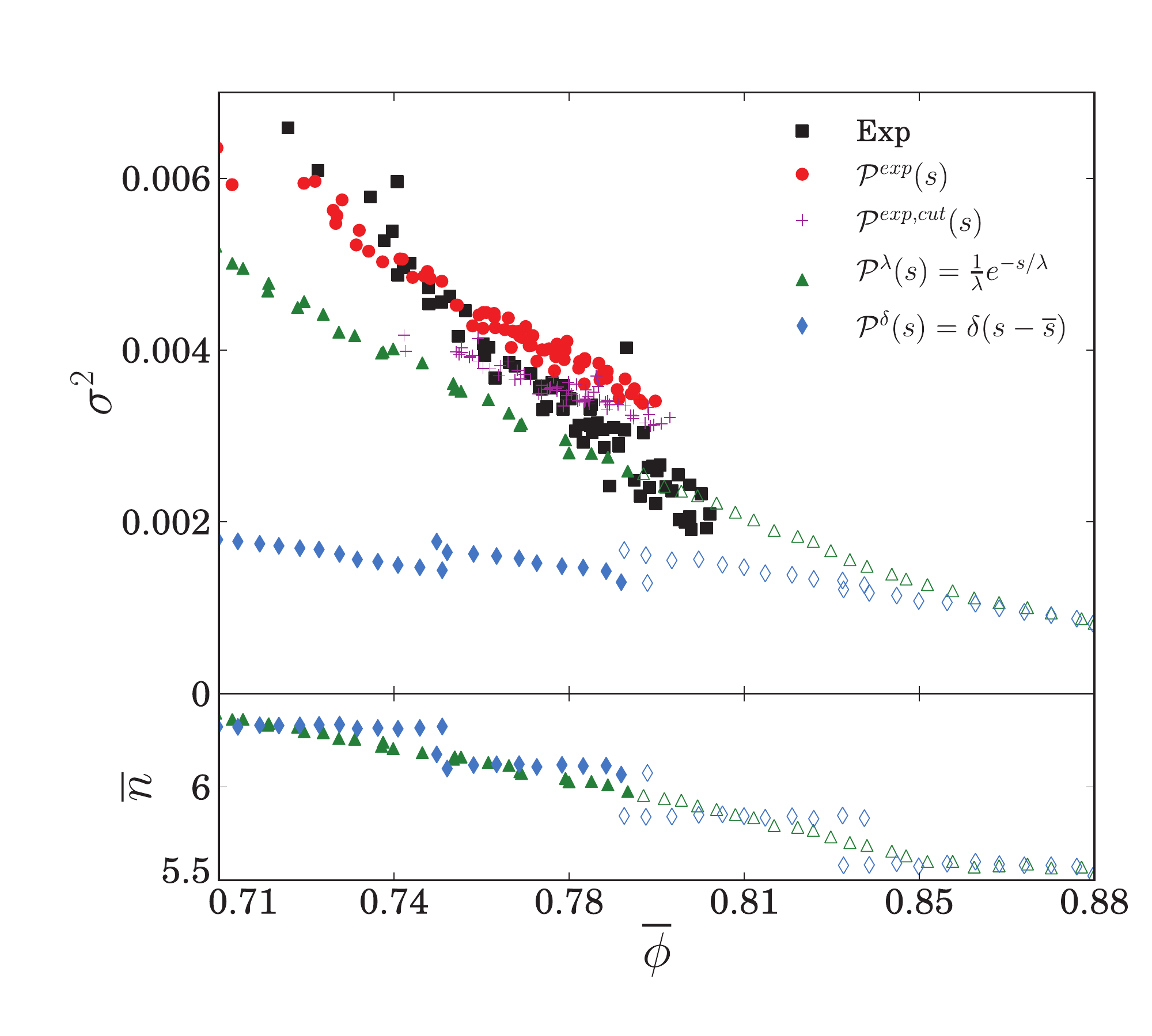}
  	\caption{Color online. (a) Mean and variance of ${\cal P}(\phi)$ for all experiments ($\blacksquare$, same data as Fig.~\ref{fig:vphi}), compared to granocentric predictions drawn from several different $s$ distributions: the experimentally-measured distribution  ${\cal P}^{exp} (s)$ ($\bullet$), the $s < r_S$ portion of the experimentally-measured distribution ${\cal P}^{exp,cut}(s)$  ($+$),  exponential  distribution ${\cal P}^\lambda(s) =\frac{1}{ \lambda} e^{-s/ \lambda} $ ($\blacktriangle$) where $\lambda = \overline{s}_{exp}$, and ${\cal P}^\delta(s) = \delta(s-\overline{s}_{exp})$ ($\blacklozenge$). The filled markers represent distribution in the experimentally observed range of $0.154 < \overline{s}_{exp} / r_S  < 0.339$.  The open markers, $\lozenge$ and $\triangle$ are ${\cal P}^\delta(\phi)$ and ${\cal P}^\lambda(\phi)$ for $\overline{s}<\overline{s}_{exp}$. (b) The mean number of neighbors versus $\overline{\phi}$ for ${\cal P}^\lambda(s)$ and ${\cal P}^\delta(s)$, with the same markers as in (a). }
	\label{fig:granocentric}
\end{figure} 

Using ${\cal P}^{exp}(s)$ as the input to the granocentric model, as described above, provides a prediction for ${\cal P}^{g}(\phi)$ which is in quantitative agreement with the experimental results. This comparison is shown in Fig.~\ref{fig:pdf-phi}ab, for the same two runs as in Fig.~\ref{fig:delta}b. As expected, the width of the distribution narrows with increasing $\phibar$ towards the random close packed limit.  In Fig~\ref{fig:pdf-phi}, all three panels compare the ${\cal P}(\phi)$ observed in the experiment and simulation with the granocentric prediction ${\cal P}^g(\phi)$. In Fig.~\ref{fig:pdf-phi}abc, we observe quantitative agreement in the peak and shape of ${\cal P}(\phi)$. However, for denser packings ($\phibar \gtrsim 0.75$), the model is systematically high on the low-$\phi$ side of ${\cal P}(\phi)$.

This comparison is performed on a small number of particles over many configurations. To test the model for a larger number of particles in a single configuration with a different preparation protocol, we use the {\tt YADE} discrete element model \cite{Kozicki-2008-YADE, Kozicki-2008-OSS} with $10^4$ particles, the same bidispersity as the experiment, and inter-particle friction coefficient $\mu_2$. The simulation is similar to the Lubachevsky-Stillinger  algorithm \cite{Lubachevsky-1990-GPR}, where each particle starts as a point with $r=0$ and is grown linearly in time proportional to the desired radius with a damping coefficient of $\eta = 0.3$. For the purposes of comparison with experimental data (which is not in a jammed configuration) we end the inflation algorithm when the volume fraction reaches $\phibar = 0.742$. Using the particle positions and sizes, we repeat the same $\phi$ and $s$ measurements as for the experiments. 

The measured ${\cal P}^{sim}(s)$ is shown in Fig~\ref{fig:pdf-phi}d and exhibits a similar shape to the ${\cal P}^{exp}(s)$ shown in Fig.~\ref{fig:delta}, with an exponential decay for $r > r_S$. Note that the peak in ${\cal P}^{sim}(s)$ is located at $s=0$, whereas the peak in ${\cal P}^{exp}(s)$ is finite (but small). The difference may be due to the differing preparation protocols. In the simulations, as the radii of particles grow, the overall energy in the system would increase unless kept finite by viscous damping. Thus, the dynamics of the system are slowed as the desired global $\Phi$ is reached. Since Fig.~\ref{fig:pdf-phi}d measures the final result of the simulation once all particle-overlaps are eliminated, we ensure $s \gtrsim 0$; due to the slow dynamics, most contacts remain at $s=0$. In contrast, the configurations in the experiment arise through collisions and the configuration $s=0$ is less likely.

Finally, we are able to compare how well the model can explain the experimental results shown in Fig.~\ref{fig:vphi}b. The results are shown in Fig.~\ref{fig:granocentric}a, where the black squares are all of the experimentally-measured $\sigma^2(\phibar)$ from in Fig.~\ref{fig:vphi}. We observe that of the four proposed distributions of $s$, the full ${\cal P}^{exp}(s)$ produces the best agreement and is able to capture not only the decreasing linear trend, but nearly the correct quantitative values. We find that the exponentially-rare neighbors with $s > r_S$ provide an important contribution to both the mean and variance: when they are removed from ${\cal P}^{exp}(s)$ to form ${\cal P}^{exp,cut}(s)$, the moments of ${\cal P}(\phi)$ are less-accurately reproduced.

For the two heuristic models of ${\cal P}(s)$, we are also able to produce the trend of decreasing variance, but without quantitative agreement with experiments. The exponential distribution ${\cal P}^{\lambda}(s)$ is a function of a single free parameter $\lambda=\bar s$, where the mean and variance are equal to $\bar s$ and $\bar s^2$, respectively.  By smoothly varying $\lambda$ over a range of values consistent with ${\cal P}^{exp}(s)$, we obtain systematically lower variance in $\phi$, indicating that the particular shape of ${\cal P}(s)$ is important for quantitative agreement. Even for the constant $s$  provided by ${\cal P}^{\delta}(s)$ (likely more consistent with jammed systems than the driven ones described here), a linear relationship for $\sigma^2(\phibar)$ remains, although with significantly less variance than observed in the experiments.  

Using ${\cal P}^\delta(s)$ causes $\sigma^2(\phibar)$ to become discontinuous, unlike experimental observations. This can be understood as arising from the low degree of polydispersity in the system. In Fig~\ref{fig:granocentric}b, the discontinuities in $\sigma^2(\phibar)$ and the average number of neighbors, $\overline{n}$, occur at the same values of $\phibar$, which is controlled by $\bar s$. As $\bar s$ increases ($\phibar$ decreases), the number of neighbors is approximately constant until $\Theta_{ex}$ is greater than the mean $\tilde{\Theta}$, at which point there is on average room for one more neighbor which appears as a step in both plots. For distributions of ${\cal P}(s)$ with sufficient variance, $\sigma^2(\phibar)$ and $\bar n$ are observed to be continuous.

Although the experiment does not probe $\phibar \gtrsim 0.8$ due to lengthening timescales \cite{Lechenault-2010-EGS}, we can explore ${\cal P}^\delta(s)$ and ${\cal P}^\lambda(s)$ for larger $\phibar$ (smaller $s$). These points are shown as the open markers in Fig.~\ref{fig:granocentric}.  In the limit  $\overline{s} \rightarrow 0$, $\sigma^2$ approaches the same non-zero value for both models. For comparison, a very loose arrangement of particles prepared to have $\phibar < 0.6$ by the same Lubachevsky-Stillinger algorithm used in Fig.~\ref{fig:pdf-phi}c exhibits an increase in $\sigma^2(\overline{\phi})$ through a maximum near $\overline{\phi} \approx 0.5$. Comparing the model ${\cal P}^{g}(\phi)$ with ${\cal P}^{sim}(\phi)$ for low $\phi$, the model underestimates the variance even though $\bar \phi$ is calculated with good agreement.

Thus, we find that the shape and moments of ${\cal P}(s)$ strongly affect the prediction of the local distribution of $\phi$.  Very narrow distributions such as ${\cal P}^{\delta}(s)$, or distributions which resemble only a portion of the experimentally measured distribution of $s$, failed to predict our experimentally measured $\phi$. Nonetheless for dense granular systems, all ${\cal P}(s)$ produced a monotonic decreasing relationship between $\sigma^2$ and $\overline{\phi}$, suggesting that this trend is quite robust. 

\section{Discussion} 

Independent of boundary condition and inter-particle friction $\mu$, we have found that as the mean local volume fraction $\phi$ increases, the variance monotonically decreases in our \textit{driven} granular system.  This result is reminiscent of similar trends observed in \textit{static} granular systems \cite{Nowak-1998-DFV, Schroter-2005-SSV, PicaCiamarra-2006-TSM, Aste-2008-STG, Briscoe-2008-EJM, SchroderTurk-2010-DSB}, in spite of the special nature of jammed systems. In particular, many properties of static granular systems are due to the inter-particle friction. Increasing $\mu$ in jammed systems decreases the required number of contacts from $\overline{z}=2d$ for $\mu=0$ to $\overline{z}=d+1$ for $\mu=\infty$.  In a model of local mechanical stability by \citet{Srebro-2003-RFC}, increasing $\mu$ increases both the tangential force and the maximum angle for which two particles can be mechanically stable. This would presumably lead to $\mu$-dependence in the  distribution of local $\phi$. However, in driven systems the number of configurations may well be independent of $\mu$ as there is no constraint on $\overline{z}$. In unjammed systems, the lack of $\mu$-dependence on the local $\phi$-distribution also suggests that dissipation does not play an important role in determining local $\phi$. However, prior observations in the same apparatus \cite{Lechenault-2010-EGS} indicate that driven, equilibrating subsystems in fact have ensembles for which $\Omega = \Omega(\mu)$, which appears to be in contrast with this idea. 

The lack of dependence on boundary condition is also surprising, particularly given two possible sources of anisotropy in the system: only three of the four walls provide driving in both CP and CV conditions, and the piston introduces a compressive force in the case of CP. Nonetheless, little anisotropy was observed in the angular distribution of neighbors: the bond angle order parameter $Q_6$ \cite{Torquato-2000-RCP} was indistinguishable for CP and CV systems at a the same $\phibar$. Additionally, we tested ${\cal P}(s)$ for angular dependence with respect to a reference vector in the lab frame and found that ${\cal P}(s)$ was rotationally symmetric for both boundary conditions.

In global (rather than local) measurements of system volume $V$, \citet{Schroter-2005-SSV} observed a non-monotonic relationship for static aggregates prepared through sedimentation of a fluidized bed at different flow rates. There, $\sigma^2$ fell from $\phi_{RLP}$ towards a transition point, then again rose on approach to $\phi_{RCP}$. This rise was attributed to cooperative effects within a finite number of statistically-independent regions, and may be related to the presence of spatial correlations during the approach to $\phi_{RCP}$ \cite{Torquato-2000-RCP}. \citet{Aste-2008-STG} compared the standard deviation as function of $\phi$ for a wide variety of experiments and simulations, and found non-monotonic behavior for the aggregated data across the transition from un-jammed to jammed configurations, although not for individual controlled experiments/simulations.  Fig.~\ref{fig:granocentric} suggests the interpretation that different experiments/simulations produce different ${\cal P}(s)$ and thus fall on different $\sigma^2(\phibar)$ curves, each of which has its own monotonic relation.
 
\section{Conclusion} 

We find a geometric  explanation for the universally-observed trend of decreasing fluctuations with increasing volume fraction. By generalizing the granocentric model \cite{Clusel-2009-GMR, Corwin-2010-MRP} to take the neighbor distribution ${\cal P}(s)$ as the single input to the model, we find that the variance of local $\phi$ measurements exhibits a smoothly decreasing trend as long as the first two moments of ${\cal P}(s)$ are also smoothly decreasing. Therefore, it is not surprising that $\sigma^2(\phibar)$ has been observed to decrease with $\phibar$ for a variety of preparation protocols and particle properties in both two and three dimensions. When ${\cal P}(s)$ is chosen to be the experimentally-observed distribution, the granocentric model provides quantitative agreement with the observed shape of $\cal P(\phi)$ and $\sigma^2(\phibar)$ without reference to any spatial correlations in local $\phi$ due to cooperative effects. Interestingly, this suggests that ${\cal P}(s)$ and ${\cal P}(\phi)$ both encode similar information about the distribution of free volume, with the first two moments set by $\phibar$. While ${\cal P}(\phi)$ (and the related free-volume distribution) has been well-studied in granular systems, ${\cal P}(s)$ has the advantage of being closely related to the radial distribution function, with the caveat of only considering neighbors instead of all particles.

In conclusion, the decreasing relationship $\sigma^2(\phibar)$ with respect to increasing $\phibar$ reveals the key role played by the constrained availability of angular space in the vicinity of a single particle.  This constraint holds regardless of whether the system is static or dynamic, jammed or unjammed, mono- or polydisperse, or two or three dimensional, and requires no information about the force chains or dynamics of the system. Therefore, this relationship provides a way of examining the distribution of free volume, and perhaps ultimately the full ensemble of valid configurations, in both jammed and driven granular materials.

\section{Acknowledgements}
We are grateful to Matthias Schr\"oter, Mark Shattuck, Raphael Blumenfeld, and Eric Corwin for useful discussions, and to the NSF for support under DMR-0644743.

\bibliographystyle{apsrev}

\end{document}